\documentclass[jgrga]{agutex}
\usepackage{graphicx}
\usepackage{graphics}
\usepackage{amssymb,amsmath}
\usepackage{rotating}

\def\lsim{\mathrel{\hbox{\rlap{\hbox{\lower4pt\hbox{$\sim$}}}\hbox{$<$}}}}
\def\gsim{\mathrel{\hbox{\rlap{\hbox{\lower4pt\hbox{$\sim$}}}\hbox{$>$}}}}

\authorrunninghead{Eke et al.}
\titlerunninghead{Lunar craters and neutron fluxes}

\begin{document}

\title{{The effect of craters on the lunar neutron flux}}

\authors{
V. R. Eke \altaffilmark{1},
K. E. Bower \altaffilmark{1},
S. Diserens \altaffilmark{1},
M. Ryder \altaffilmark{1},
P.E.L. Yeomans \altaffilmark{1},
L. F. A. Teodoro \altaffilmark{2},
R. C. Elphic \altaffilmark{3},
W. C. Feldman \altaffilmark{4},
B. Hermalyn \altaffilmark{5},
C. M. Lavelle \altaffilmark{6},
D. J. Lawrence \altaffilmark{6}}
\altaffiltext{1}{Institute for Computational Cosmology, Physics Department, Durham University, Science Laboratories, South Road, Durham DH1 3LE, U.K.}
\altaffiltext{2}{BAER, Planetary Systems Branch, Space Science and Astrobiology
  Division, MS: 245-3, NASA Ames Research Center, Moffett Field, CA 94035-1000, U.S.A.}
\altaffiltext{3}{Planetary Systems Branch, Space Science and Astrobiology
  Division, MS: 245-3, NASA Ames Research Center, Moffett Field, CA 94035-1000, U.S.A.}
\altaffiltext{4}{Planetary Science Institute, 1700 East Fort Lowell, Suite 106, Tucson, AZ 85719, U.S.A.}
\altaffiltext{5}{NASA Astrobiology Institute, Institute for Astronomy, University of Hawaii, 2680 Woodlawn Drive, Honolulu, HI 96822, U.S.A.}
\altaffiltext{6}{Johns Hopkins University Applied Physics Laboratory, Laurel, MD 20723, U.S.A.}

\begin{article}

\abstract{
The variation of remotely sensed neutron count rates is
measured as a function of cratercentric distance using data from the Lunar
Prospector Neutron Spectrometer. The count rate, stacked over many
craters, peaks over the crater centre, has a 
minimum near the crater rim and at larger distances it increases to
a mean value that is up to $1\%$ lower than the mean count rate observed over the
crater. 
A simple model is presented, based upon an analytical
topographical profile for the stacked craters fitted to data from the
Lunar Orbiter Laser Altimeter (LOLA). The effect of topography coupled
with neutron
beaming from the surface largely reproduces the observed count rate profiles. 
However, a model that better fits the observations can be found by
including the additional freedom to increase the neutron 
emissivity of the crater area by $\sim 0.35\%$ relative to the unperturbed surface.
It is unclear what might give rise to this effect, but it may relate
to additional surface roughness in the vicinities of craters.
The amplitude of the crater-related signal in the neutron count rate is
small, but not too small to demand consideration when inferring
water-equivalent hydrogen (WEH) weight percentages in polar
permanently shaded regions (PSRs).
If the small crater-wide count rate excess is 
concentrated into a much smaller PSR, then it can lead to a 
large bias in the inferred WEH weight percentage. For
instance, it may increase the inferred WEH for Cabeus crater at the
Moon's South Pole from $\sim 1\%$ to $\sim 4\%$.
}
\vspace{1cm}

\section{Introduction}

Cosmic ray interactions with planetary surfaces lead to nuclear
fragments being released in the regolith. The study of neutrons that
avoid nuclear recapture and subsequently escape through the surface
provides a route to determining the abundance of various nuclei near
the surface of those bodies \citep{lin61,metz90,feld91}. Of particular
interest is the epithermal neutron 
flux (energies in the range $0.3$ eV$<E<0.5$ MeV), because of
its sensitive dependence on the hydrogen abundance in the top $\sim
70$ cm of regolith \citep{feld00b}. The first
experiment to search for lunar hydrogen in this way was the Lunar
Prospector Neutron Spectrometer \citep[LPNS,][]{feld04}. \cite{feld98}
found that there 
were polar dips in the epithermal neutron count rate, implying the
existence of polar near-surface hydrogen. Furthermore, the lack of a
corresponding feature in the fast neutrons with energies exceeding
$0.5$ MeV \citep{feld98} or the
thermal neutrons with $E<0.3$ eV \citep{law06} suggested that any
hydrogen-rich layer of material should be buried beneath $5-10$ cm of
hydrogen-poor material.

The omni-directional LPNS, when orbiting at $30$ km, had a spatial
footprint with a full-width half-maximum (FWHM) on the lunar surface
of $45$ km \citep{maur04}. 
In order to suppress statistical noise, \cite{feld98} and
\cite{feld01} binned the data into $\sim 60$ km $\times 60$ km pixels.
This is large relative to the sizes of most permanently shaded
regions. However,
\cite{eke09} showed, by stacking data and using a pixon-based image
reconstruction technique to improve the spatial resolution while
suppressing the noise, that these count 
rate dips could, in a statistical sense, be associated with the
permanently shaded regions. This result was confirmed by \cite{luis10}
using a more accurate set of permanently shaded regions defined by
the SELENE laser altimeter \citep{noda08}.
The count rate dip inferred for Cabeus crater
corresponded to a $(1\pm0.3)~{\rm wt}\%$ water equivalent hydrogen
(WEH) according to the regolith composition model of \cite{law06}, which has a
semi-infinite layer of ferroan anorthosite (FAN)-type soil with
varying amounts of H$_2$O. 

Another experiment, the Lunar Exploration Neutron Detector (LEND),
contained sensors called the Collimated Sensor for EpiThermal Neutrons
(CSETN) and the Sensor for EpiThermal Neutrons
\citep[SETN,][]{mit10a}. These mapped the Moon from the Lunar Reconnaissance
Orbiter at an altitude of $50$\, km, $\sim 20$ km above the orbit of Lunar
Prospector. Thus, one should not expect the SETN instrument to provide
competitive results relative to the LPNS. Furthermore, comprehensive
analyses of the data returned from the CSETN have demonstrated that
the collimator did not perform well enough to fulfil its mission
objectives \citep{law11b,miller12}, with the vast majority of lunar neutrons
being uncollimated \citep{eke12} and an effective FWHM much larger
than that of the omni-directional LPNS \citep{luis14}.
In view of the difficulties associated with the interpretation of this
data set, these data will be considered only briefly in this paper.

The Lunar Crater Observation and Sensing Satellite (LCROSS) impacted
into Cabeus crater in 2009 and the resulting ejecta plume was analysed to
give a value of $(5.6\pm2.9)~{\rm wt}\%$ WEH \citep{cola10}. While
statistically consistent with the LPNS result, the most probable value
is over five times the LPNS-inferred value. The reanalysis of the
LCROSS data by \cite{stry13}, which gave $(6.3\pm1.6)~{\rm wt}\%$ WEH,
is inconsistent with the LPNS result.
These comparisons would be affected if the
hydrogen detected by the LPNS was not uniformly spread across the
surface within the large resolution element, which is approximately
1000 times as long as the crater produced by the LCROSS impact
\citep{schul10}. 
The LPNS and LCROSS results sample somewhat different depths into the
regolith. Thus, any variation in hydrogen content with depth could
also lead to a difference between the hydrogen abundances
inferred from the two separate methods. One assumption that is
implicit in the studies of craters using the LPNS data is that neutron
count rates are not explicitly affected by the surface
topography. A new model will be presented in this paper to quantify
the topographical effect from craters on the neutron count rate.

The Chandrayaan-1 M$^3$ results interpreted as implying a particular
excess of water or hydroxyl molecules in Goldschmidt crater
\citep{pie09} prompted \cite{law11} to 
re-examine LPNS data in this region in the context of a two-layer
regolith model, with the surface layer being hydrogen-rich. This
contrasted with previous Monte Carlo modelling of the lunar regolith,
where the hydrogen had been buried under a dry layer of regolith \citep{law06}.
After removal of the trends caused by bulk composition,
the thermal and epithermal data in the vicinity of Goldschmidt crater
were compared with the models to
investigate the sensitivity of neutron measurements to the depth
distribution of hydrogen. \cite{law11} concluded that it was necessary to
understand more about systematic variations at the $1-3\%$ level before
definitive conclusions could be reached. If crater topography does
provide small systematic variations in neutron count rates then
it needs to be understood in order to progress.

When studying the LPNS count rate, \cite{feld01} found that `local maxima 
overlay the floors of large, flat-bottomed craters'. They did not
speculate as to what this implied, but the possibility of a
topographical effect on the measured neutron count rate is one that could
create a systematic bias in the values of WEH inferred above
permanently shaded polar craters. To date there has been no
systematic, quantitative study of the imprint of topographical 
features on the detected orbital neutron count rate. It is important to
quantify the impact of topography on the emitted lunar neutron flux because
many of the results from the LPNS involve
small changes in count rates measured over craters.

Section~\ref{sec:data} contains a description of the neutron and topography
data being used. Fits to the crater average topography are given in
Section~\ref{sec:lola}. The variation of neutron count rate as a
function of distance to the crater centre is shown in 
Section~\ref{sec:res}, for a variety of different subsets of craters.
In Section~\ref{sec:mod}, a simple model is presented for how the neutron count rate
changes as a function of detector distance from the crater
centre. This model is confronted with the data, and the implications
for our understanding of the regolith are
investigated. Section~\ref{sec:imp} discusses the implications of this
work for quantitative estimates of cold-trapped hydrogen, and
conclusions are drawn in Section~\ref{sec:conc}. 

\section{Data}\label{sec:data}

Maps of the lunar neutron count rate, a set of predetermined lunar craters and a
digital elevation map are necessary to calculate the neutron count
rate profiles near craters. The data sets to be used here, which are
all available from the Geosciences Node of NASA's Planetary Data
System (PDS\footnote{http://pds-geosciences.wustl.edu}), are
described in this section.

\subsection{Lunar Prospector neutron data}\label{ssec:lpns}

The Lunar Prospector spacecraft spent one year at $100$ km altitude, then seven
months at $40-30$ km. PDS time series data from the thermal,
epithermal and fast 
neutron detectors, processed as described by \cite{maur04}, are used in
this study, with the focus mainly on the low-altitude subset. Some
results from the high-altitude period will also be shown for
comparison, but the default choice is to consider only data for which
Lunar Prospector was at an altitude less than $45$ km. 
Using different energy neutrons is desirable because of their
differing responses to changes in regolith composition.
Also, the thermal neutrons
probe further into the regolith than the epithermals, whereas the fast
neutrons typically sample nearer to the surface than the epithermals.

\subsection{Lunar Exploration Neutron Detector data}\label{ssec:lend}

Data from the first 15 months of the mapping orbits are used for both
the LEND SETN and CSETN detectors, to compare with the results from
the LPNS. For the CSETN measurements the 
background due to cosmic rays striking the spacecraft itself is
removed statistically following the procedure described by
\cite{eke12}. The remaining count rate is comprised of two distinct lunar
components, where the detected neutrons originate either from within
or outside the collimator's geometrical field-of-view. 
One cannot determine from which component individual neutrons originate.

\subsection{Crater list and topographical data}\label{ssec:crat}

The list of craters produced by \cite{head10} from the Lunar Orbiter Laser
Altimeter (LOLA) topographical
data is used. This consists of 5185 craters with radii of at least
$10$ km, distributed over the entire lunar surface. In this study,
various different selections of craters are made, based on the radii,
$r_c$, and the central locations given in this list. The variable $r$
will be used here to represent arclengths along an
unperturbed spherical surface, whereas the variable $x$ represents the
distance from the symmetry axis ($z$) of a crater. Thus, the measured
crater diameters are really $2x_{\rm c}$ in this nomenclature. $x$ and $r$
are related via
\begin{equation}
x=r_{\rm m} \sin\left(\frac{r}{r_{\rm m}}\right),
\label{xofr}
\end{equation}
where $r_{\rm m}=1737.4$ km is the lunar radius. This equation implies
that $x_{\rm c}$ will be within $0.1\%$ of $r_c$ for crater radii less than
$100$ km, so the variables $x_{\rm c}$ and $r_c$ will be assumed to be equal
for the rest of this study. The angle subtended at the lunar centre by
the crater radius is
\begin{equation}
\theta_c=\sin^{-1}\left(\frac{x_{\rm c}}{r_{\rm m}}\right).
\end{equation}

The global topographic map from LOLA \citep{smith10} with
$(1/64)^\circ$ resolution is used to measure the crater topographical
profiles. This corresponds to $\sim 0.5$ km resolution at the equator,
which is more than sufficient for the
approximate modelling of crater topography as a function of crater
radius that is necessary for the neutron count rate model presented 
in Section~\ref{sec:mod}.

The epithermal neutron count rate measured by the omnidirectional LP detector
changes by approximately $10\%$ across the whole Moon. This variation
is dominated by known changes in regolith composition. Any systematic
topographical effects are expected to be at the level of $\sim 1\%$, as
noted by \cite{feld01}. 
This anticipated variation is sufficiently small that it is
necessary to stack together craters of similar size in order to 
reduce the statistical uncertainties. 
In addition, the stacking averages away azimuthal anisotropy that
exists in the crater sample, making radial profiles an appropriate way to
represent the results.
To produce a more homogeneous set
of craters to stack together, both in terms of topography and
composition, only craters in `highland' regions will be considered in
this study. This means only craters on the far side of the
Moon and with latitudes greater than $-20^\circ$ will be included in
the stacking procedure. These cuts leave just 2216 craters.
This choice is important for some of the results presented later
involving thermal and fast neutrons, which are both more sensitive
than epithermal neutron fluxes to iron and titanium abundances.

\section{Crater topography}\label{sec:lola}

The model for the topographical effect on the neutron count rate described
in Section~\ref{sec:mod} needs to assume a 
particular crater profile. This section first describes the functional
form of the assumed crater profile, then measures it using the LOLA
digital elevation map by stacking radial profiles for craters in the chosen range of
sizes.

\subsection{Model crater profile}\label{ssec:lolamod}

\begin{table*}
\caption{Least-squares parameter values for the crater topography model
fitted to the stacked LOLA data as a function of crater
radius.\tablenotemark{a}
}
\centering
\begin{tabular}{c|c|c|c|c|c|c}\hline
Crater radius, $r_c/$km & Number of craters & Depth, $d$/km & Infill radius, $x_{\rm i}/x_{\rm c}$ &
Edge of outer slope, $x_{\rm e}/x_{\rm c}$ & Uplift, $u$/km \\ \hline
$10-20$ & 1264 & $2.00$ & $0.30$ & $2.1$ & $0.45$\\
$20-30$ & 482 & $2.50$ & $0.42$ & $2.1$ & $0.60$\\
$30-40$ & 219 & $2.75$ & $0.48$ & $2.0$ & $0.70$\\
$40-50$ & 115 & $3.25$ & $0.56$ & $2.2$ & $0.85$\\
$50-60$ & 47 & $3.70$ & $0.66$ & $2.1$ & $1.15$\\
$60-80$ & 42 & $3.40$ & $0.70$ & $1.8$ & $1.00$\\
$40-50$ deep & 57 & $4.05$ & $0.48$ & $2.2$ & $1.15$\\
$40-50$ shallow & 58 & $3.05$ & $0.69$ & $2.1$ & $0.65$\\
\hline
\end{tabular}
\tablenotetext{a}{The depth, fractional infill radius, fractional edge of
outer slope and uplift are found on grids with resolution $0.05$ km,
$0.01$, $0.01$, and $0.05$ km respectively, which are larger than the
statistical uncertainties on these parameters. The midpoint of the crater
radius range is used to calculate the parameters for each stacked profile.}
\label{tab:modpar}
\end{table*}

\begin{figure}
\begin{center}
\includegraphics[trim=2cm 7cm 1cm 3cm,clip=true,width=0.95\columnwidth]{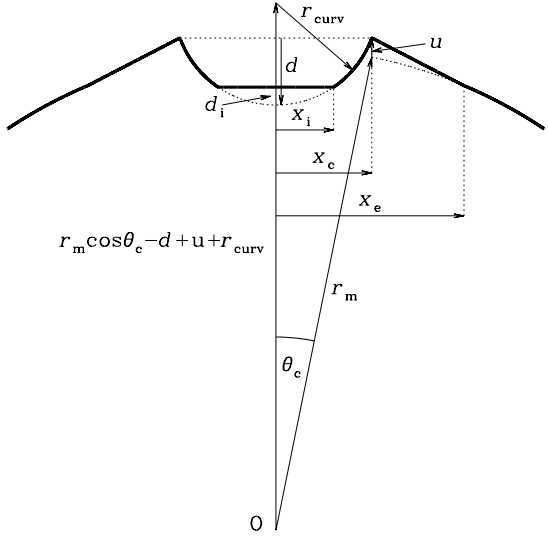}
\end{center}
\caption{Model crater profile (bold line) that is fitted to the LOLA
topographical data for the selected craters. The model takes into
account a flat infill region at radii less than $x_{\rm i}$, a spherical cap
depression out to $x_{\rm c}$ and an outer slope, uplifted at the crater rim by an amount $u$,
that returns to the unperturbed surface at $x_{\rm e}$. The centre of the Moon is used as
the origin of the coordinate system.
}
\label{fig:crater}
\end{figure}

Rather than having a general topography, the model craters in
Section~\ref{sec:mod} are considered to have azimuthally symmetric
profiles of a kind shown in Figure~\ref{fig:crater}. 
These consist of a spherical cap depression of depth $d$
measured down from the $x-y$ plane containing the crater rim, with a central,
flat ($dz/dx=0$) infill region extending out to a radius $x_{\rm i}$, and
with a maximum depth, at $x=0$ of $d_{\rm i}$, where the i subscript
refers to the infill region. The radius of
curvature for the spherical cap part of the crater is then
\begin{equation}
r_{\rm curv}=\frac{x_{\rm c}^2+d^2}{2d},
\end{equation}
and the maximum infill depth, measured from the base of the spherical
cap to the infill surface, is given by
\begin{equation}
d_i=r_{\rm curv}-\sqrt{r_{\rm curv}^2-x_{\rm i}^2}.
\end{equation}
This crater is
uplifted parallel to the crater axis (the $z$ direction) by a distance
$u$, with an outer slope of constant gradient, $dz/dx\equiv g$, back to the
unperturbed surface at a perpendicular distance of $x_{\rm e}$ from the
symmetry axis of the crater. Defining
\begin{equation}
\theta_e=\sin^{-1}\left(\frac{x_{\rm e}}{r_{\rm m}}\right),
\end{equation}
the gradient of the outer slope is given by
\begin{equation}
g=\frac{r_{\rm m}\cos\theta_e-(r_{\rm m}\cos\theta_c+u)}{x_{\rm e}-x_{\rm c}}.
\label{grad}
\end{equation}

The height of the crater relative to an unperturbed surface can be
found using
\begin{equation}
h(x)=\sqrt{x^2+z(x)^2}-r_{\rm m},
\label{hofx}
\end{equation}
where $z$ is defined as zero at the lunar centre.
For $x\geq x_{\rm e}$, $h=0$ for the model crater. At $x=0$, 
\begin{equation}
z(x=0)\equiv z_0=r_{\rm m}\cos\theta_c-d+u+d_i.
\end{equation}

At radii where the presence
of the crater perturbs the surface, the height can be inferred using
the following expression for $z$:
\begin{equation}
z(x)=\left\{  \begin{array}{ll}
             z_0 & \mbox{if $x\leq x_{\rm i}$~;} \\
             z_0-d_i+r_{\rm curv}-\sqrt{r_{\rm curv}^2-x^2} & \mbox{if $x_{\rm i}\leq x \leq x_{\rm c}$~;} \\
             z_c+g(x-x_{\rm c}) & \mbox{if $x_{\rm c}\leq x \leq x_{\rm e}$~.}
             \end{array}
             \right.
\label{zofx}
\end{equation}
$z_c$ represents 
$z(x_{\rm c})\equiv z_0-d_i+r_{\rm curv}-\sqrt{r_{\rm curv}^2-x_{\rm c}^2}$. 

A least-squares minimisation was performed to find the best-fitting sets of
the parameters $[d,x_{\rm i},u,x_{\rm e}]$ for the subsets of craters
of different radius. For all subsets of craters with radii of at least
$20$ km, the region at $x<0.2x_{\rm c}$ was excluded from the fit,
because a central peak, not included in the model, often exists. The
midpoint of the crater radius bin was chosen for the $x_{\rm c}$ of
the model to be fitted to the stacked profile. 

\subsection{Crater topography fits}\label{ssec:loladat}

For each crater, digital elevation map measurements within $3r_c$ were
used to construct the relative height profile as a function
of $r/r_c$, where $r$ represents the arc length from the crater centre
to the spacecraft nadir. The zero of height for each crater is defined 
as the mean height in the range $2.5<r/r_c<3$.
Each measurement provides an estimate of the relative height at
its $r/r_c$. The statistical uncertainty on the estimated
mean height is just the square root of the ratio of the variance of the
individual measurements within a given bin in $r/r_c$ to the number of
observations in that bin.
The craters were stacked by crater radius, because the typical crater
shape varies systematically with crater radius. 

The crater set with $40<r_c/$km$<50$ was further subdivided by depth
to see how this affected the neutron count rate profiles.
To split the crater subset by crater depth, in order to investigate
the effect on the neutron profile, the depth of each crater is defined
as the difference between the average heights in the radial ranges
$(0.95-1)r_c$ and $(0.2-0.3)r_c$. The central region is once again
avoided to reduce any systematic effect due to central peaks.
While this statistic might, in some instances, reflect 
subcraters rather than the larger scale topography, it
at least serves as a simple way to separate deep and shallow 
craters with the same radius.

\begin{figure}
\begin{center}
\includegraphics[trim=1cm 6cm 1cm 3cm,clip=true,width=0.95\columnwidth]{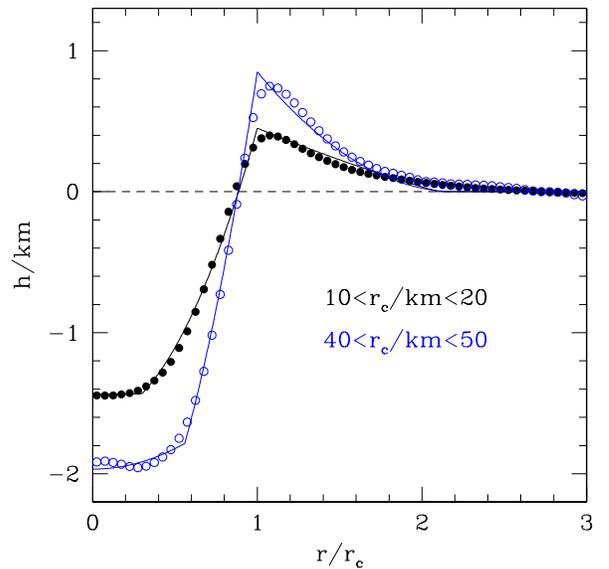}
\end{center}
\caption{
Mean radial LOLA topographical profiles for craters with radii in the
ranges $10-20$ km (black filled circles) and $40-50$ km (blue open
circles). Statistical
errors on the mean profiles are smaller than the symbol
sizes. Solid lines show the least-squares model fits to the
data sets as detailed in equations~(\ref{xofr}-\ref{zofx}).
}\label{fig:topofit}
\end{figure}

Figure~\ref{fig:topofit} shows the azimuthally averaged mean radial
topographical profiles for craters of different sizes, as measured
using craters from the \cite{head10} list and LOLA topographical
data. Statistical errors on the mean profiles are smaller than the symbol
sizes. It is apparent that for the larger craters there are central
peaks that are not included in the model profile, as described in the
previous section.
\begin{figure*}
\begin{center}
\includegraphics[trim=1cm 16cm 1cm 4cm,clip=true,width=1.95\columnwidth]{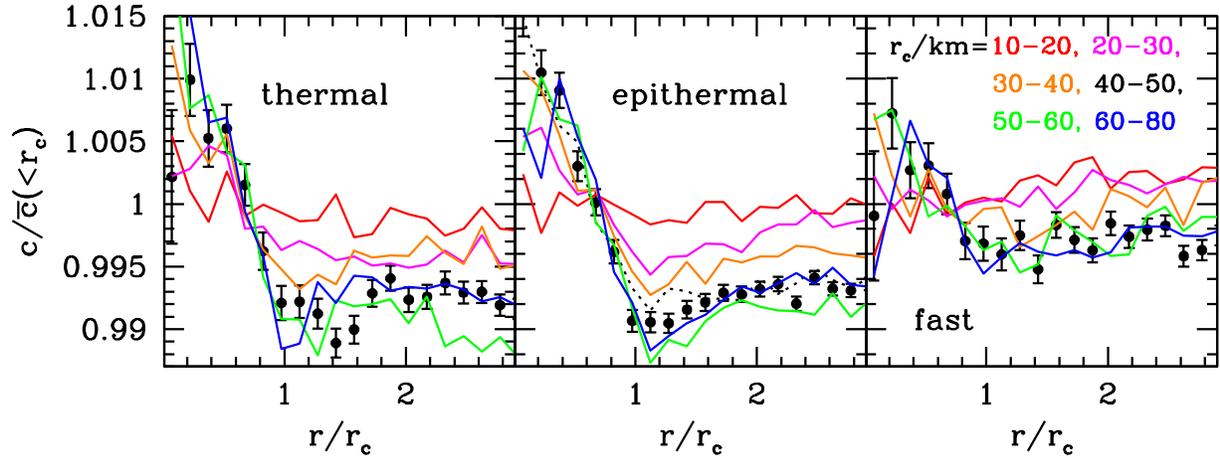}
\end{center}
\caption{
Stacked, normalised radial LP neutron mean count rate profiles for
different radius 
craters. Thermal, epithermal and fast neutron results are shown in the
left, centre and right panels respectively. The different colours
correspond to craters in the radius ranges $10-20$ km (red), $20-30$
km (magenta), $30-40$ km (orange), $40-50$ km (black points), $50-60$
km (green) and $60-80$ km (blue). A black dotted line represents the
epithermal neutron profile for $40-50$ km craters when the time series is
sampled in $32$ s observations, like the fast neutron data set, rather
than $8$ s like the thermal and epithermal time series.
}\label{fig:neurc}
\end{figure*}
The best-fitting models are also shown in Figure~\ref{fig:topofit}, from
which it can be seen that the model becomes increasingly inappropriate
for larger craters. A flat,
$dz/dx=0$ central region does not translate to a constant height
relative to the unperturbed spherical surface, which would provide a
better fit to the $r_c>70$ km craters. Also, the constant $dz/dx$ outer
slope, at large distances, can lead to $h<0$ on the outer uplifted
slope; a feature not present in the observations. Despite these
shortcomings in the model, it does capture the main features
present in the measured average topographical profiles, and the extent
to which the model is inadequate is not quantitatively significant for
the neutron count rate results in subsequent sections.

Table~\ref{tab:modpar} lists the best-fitting model parameters for a
set of different crater size ranges. These values are used in
Section~\ref{sec:mod} for the model predicting the topographical
effect on the neutron count rate profiles observed by the orbiting
detector. The depth parameter, $d$, only represents the depth of
the crater when there is no infill so, as can be seen by
comparing the values in Table~\ref{tab:modpar} with the data in
Figure~\ref{fig:topofit}, the actual crater depths
from rim to minimum are typically much smaller than $d$. This is
particularly true for the larger craters, where the best-fitting
infill region extends to a larger fraction of the crater radius.

\section{Neutron count rate profiles}\label{sec:res}

For each crater, time series observations within $3r_c$ were
used to construct relative count rate profiles, where the relative
count rate is defined for each 
crater by dividing each time series measurement by the mean count rate
within $r_c$ of the crater centre. Each time
series observation provides an estimate of the relative count rate at
a given $r/r_c$ and these are stacked together for different crater subsets. 

\subsection{Radial count rate variations}\label{ssec:neures}

The results in this subsection show how the neutron count rate varies
as a function of sub-detector point distance to the crater
centre. Stacked subsets of similar-sized highland craters are used,
as are data for different neutron energy ranges.

Figure~\ref{fig:neurc} shows the mean stacked, count
rate ($c$) profiles, with observations from each contributing crater
normalised by the mean count rate measured from positions over that
crater, $\bar{c}(<r_c)$. As the craters are very well sampled, had the
count rates instead been normalised with respect to the count rate
at $r/r_c\sim 3$ the stacked profiles would only change by a
radius-independent, vertical shift. The radii are normalised by the relevant 
crater radius. Points and errors on the mean profiles are shown only for the
$40<r_c/$km$<50$ case for clarity, but are of similar size for the
other crater subsets. For both the thermal and epithermal profiles, a
central $\sim 1\%$ enhancement in the neutron count rate is seen, with
the count rate outside the crater being $\sim 0.7\%$ lower than the
mean count rate measured over the crater. These features are
about twice as pronounced as those in the corresponding fast neutron
profiles, and are common to all crater samples with $r_c>40$ km. For
smaller crater sizes, the features in the profiles
decrease in amplitude. Given that the FWHM of the LP neutron
detectors is approximately $45$ km at an altitude of $30$ km, one
should expect that any features on smaller scales will be washed out.
Also, if all craters have their radii either
overestimated or underestimated by $5\%$ then the changes in the
neutron count rate profiles are lower than $0.1\%$, so the results are
robust to this level of systematic uncertainty in the crater radius
determination. 

One reason why the fast neutron profile might be less variable than
those in the lower energy ranges is that the temporal sampling is
lower, with $32$ s observations, rather than $8$ s. At an altitude of
$30$ km, LP travelled $\sim 50$ km during $32$ s. The effect of the
resultant blurring of the profile can be estimated by degrading the
sampling of the epithermal neutron time series. This is illustrated for
the $40<r_c/$km$<50$ craters by the dotted line in the central panel
of Fig.~\ref{fig:neurc}. For craters that are at least this large, the
different 
sampling has only a small effect. However, for smaller craters, where
the distance travelled during an individual observation corresponds to
a larger $r/r_c$, the suppression of features in the normalised count rate
profile will be larger. Fig.~\ref{fig:neurc}
suggests that any features in the fast neutron profile for craters
with $r_c<40$ km would be small anyway.

\begin{figure}
\begin{center}
\includegraphics[trim=0cm 6cm 1cm 3cm,clip=true,width=0.95\columnwidth]{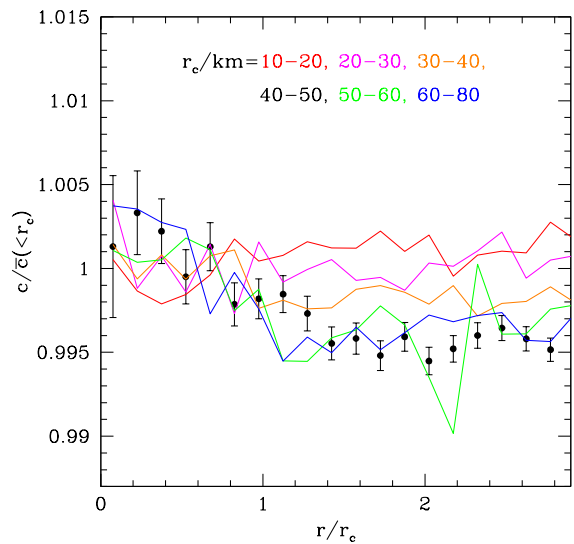}
\end{center}
\caption{
Stacked, normalised radial LEND SETN neutron count rate
profiles for observations at altitudes less than $60$\,km. The different colours 
correspond to different sized craters, as described in the caption for 
Figure~\ref{fig:neurc}.
}\label{fig:setn}
\end{figure}

Lunar Prospector neutron count rates are evidently affected, at the $\sim
\pm 1\%$ level, by the detector position relative to craters on the
surface, provided that the detector footprint is small enough to allow
it to `see' the craters. At this point, it is worth briefly
considering the count rate profiles produced by the LEND SETN and
CSETN. The SETN is an ``omni-directional'' detector, albeit strapped
to the side of a ``collimator'', so in practice it has an
energy-dependent anisotropic footprint. Given that it is viewing the
surface from $\sim 50$ km altitude, the features
seen by the LPNS should be stronger than those recovered by the
SETN. This is evident in Figure~\ref{fig:setn}, which shows the count
rate profiles measured by the SETN for a range of different crater
sizes. The features, while still significant, have been washed out,
typically decreasing the amplitude of the central peak by a factor of a few.

\begin{figure}
\begin{center}
\includegraphics[trim=0cm 6cm 1cm 3cm,clip=true,width=0.95\columnwidth]{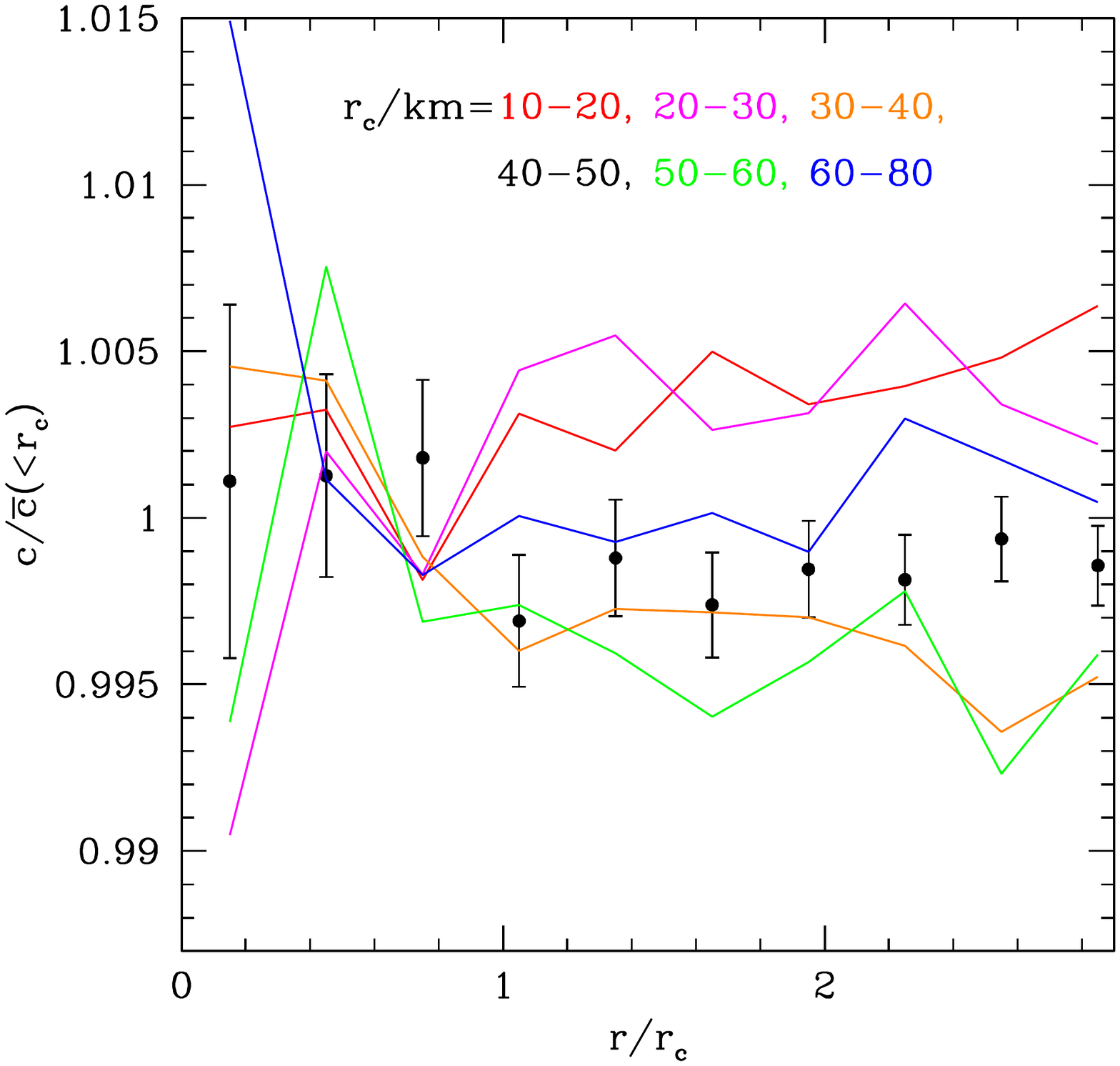}
\end{center}
\caption{
Stacked, normalised radial LEND CSETN neutron count rate
profiles for observations at altitudes less than $60$\,km. The different colours 
correspond to different sized craters, as described in the caption for 
Figure~\ref{fig:neurc}.
}\label{fig:csetn}
\end{figure}

Figure~\ref{fig:csetn} shows the spacecraft background-corrected CSETN
count rate profiles. After correction, the count rates are typically
only $\sim 2$ per second, hence the large statistical
uncertainties. However, unlike the SETN profiles that show similar
trends with crater size to the LPNS results, the CSETN profiles show
no obvious trends or significant central bumps in the count rate. This
is entirely consistent with the large CSETN detector footprint
inferred by \cite{luis14}.


Having determined that the LPNS count rate varies systematically with
distance from crater centres, the question becomes what
is responsible for this? One uninteresting possibility can be
immediately discounted by recalculating the count rate profiles using
the raw LPNS data. The features are similarly present in the raw data,
implying that the data reduction process did not create them and they
do reflect something to do with the lunar surface. 
Compositional variation would not create almost identical features
in the thermal and epithermal count rate profiles. Also, if mafic
and magnesian central peaks \citep{cahill09} were having an important
impact on these profiles, then the thermal and fast neutron profiles
should be anticorrelated, whereas they show qualitatively similar behaviour.
The 
possibility that these profiles are the result of the geometrical
configuration will be considered in the following section.


\section{A Simple Geometrical Model}\label{sec:mod}

\begin{figure*}
\begin{center}
\includegraphics[trim=0cm 6cm 0cm 3cm,clip=true,width=1.95\columnwidth]{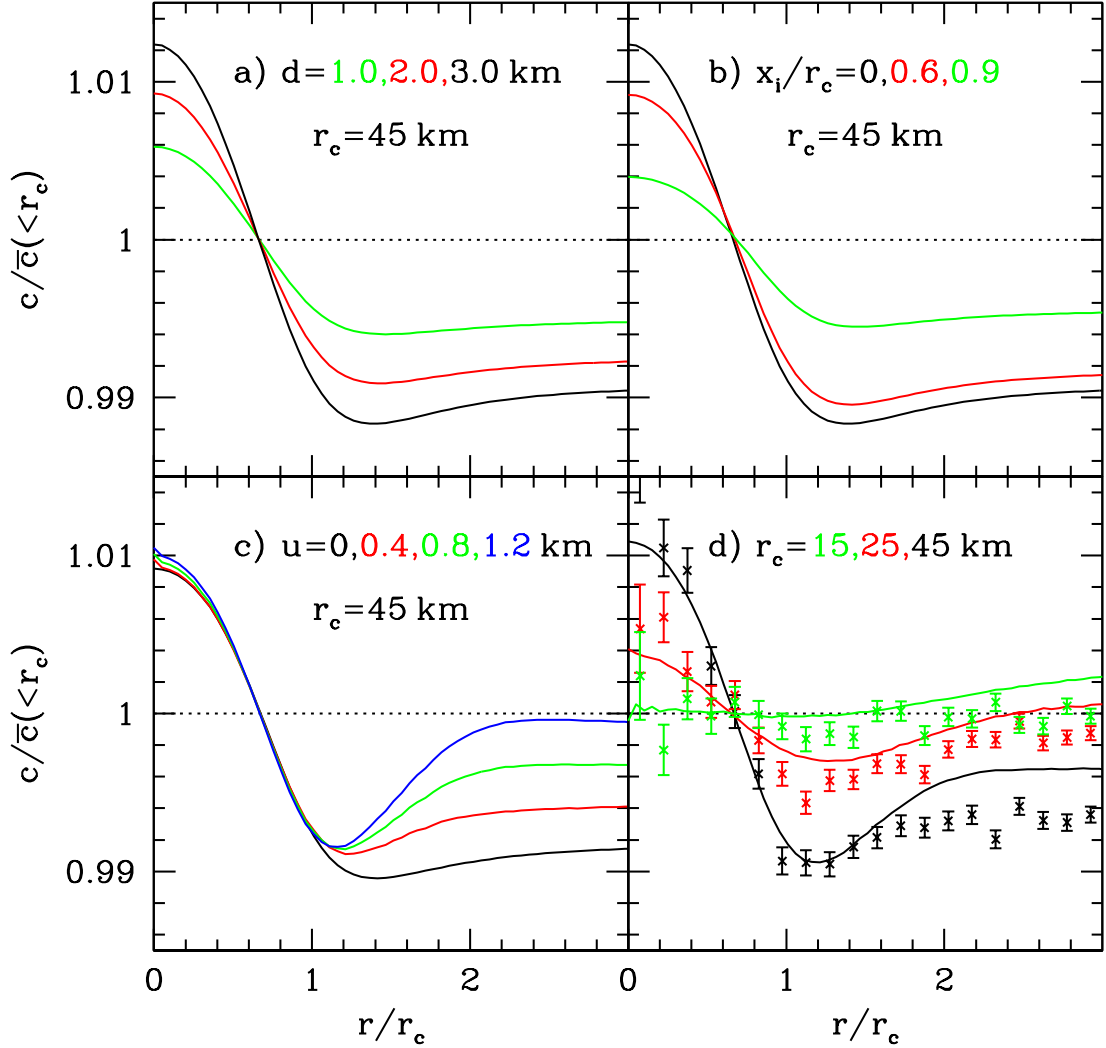}
\end{center}
\caption{
Variation of the model count rate profile normalised to the mean count
rate over the crater. Panel (a) shows the variation with crater
depth, $d$, for a $45$\,km radius crater, with no 
infill or uplift, and the detector placed at an altitude of $30$\,km. 
These model craters are just spherical cap depressions. Panel (b)
fixes $d=3.0$\,km and shows the effect of adding a flat central infill
region out to a fraction $x_{\rm i}/r_c$ of the crater radius. Fixing $x_{\rm i}=0.6\,r_c$
and $x_{\rm e}=2.0\,r_c$,
panel (c) illustrates the variation of the count rate profile with
crater uplift. Finally, panel (d) shows how the model profile varies
with crater radius. At each different radius, the crater shapes are
defined using the appropriate parameters in
Table~\ref{tab:modpar}. The data points show the LPNS count rate
profiles for the corresponding crater stacks.
}\label{fig:mod4}
\end{figure*}

A model describing how topography affects the detected neutron flux is
outlined in this section, in order to determine if this alone can
explain the neutron count rate profile over craters. The
predictions from this model are compared with the LPNS count rate
profiles and the implications of this comparison are then discussed.

\subsection{The model}\label{ssec:mod}

The flux measured a distance $r$ away from a patch of surface area
d$A$ emitting neutrons
at a rate $f_0$ per unit area, with a detector an angle $\theta$ away
from the surface normal can be written as
\begin{equation}
{\rm d} f(r,\theta)=\frac{(2+\alpha)f_0}{2\pi
  r^2}\cos^{1+\alpha}\theta \,{\rm d}A,
\label{modf}
\end{equation}
where $\alpha$ represents the effective beaming of the neutrons out of
the surface resulting from the increase in neutron number density
with depth in the top $\sim$ mean free path in the
regolith \citep{mck06}. $\alpha\approx0.5$ provides a good match to
the Monte Carlo neutron transport flat surface models of \cite{law06} for
the range of neutron energies detected by the LPNS.

The model for the total flux received by the detector involves
integrating equation~(\ref{modf}) over the lunar surface that is visible
from the detector, assuming that the flux from a particular
piece of surface is proportional to the incident cosmic ray flux. One
complication is that when the surface includes concave 
craters, their walls can act to block parts of the crater interior
from the detector's view. If crater uplift is included then this
effect can also extend to the exterior of the crater. The model
accounts for this but does not, by default, allow neutrons emitted
from the crater and impinging on the crater wall to be reemitted. This
assumption will be considered further in section~\ref{ssec:imp}.

In practice, the flux calculation can be more efficiently performed by
partitioning the surface into different zones and using symmetries in
the problem to avoid needing to do a two-dimensional numerical
integration over the full visible surface. These zones are:
\begin{enumerate}
\item the central infilled region of the crater, $(x\le x_{\rm i})$,\\
\item the constant radius of curvature crater walls, $(x_{\rm i}\le x\le x_{\rm c})$,\\
\item the outer uplifted slope, $(x_{\rm c}\le x\le x_{\rm e})$,\\
\item the unperturbed surface beyond the outer uplifted slope, $(x_{\rm e}\le x)$.\\
\end{enumerate}
For the more interested reader, Appendix~\ref{app:mod} 
contains specific details of the calculations involved.

\subsection{Predictions of the model}\label{ssec:modres}

Figure~\ref{fig:mod4} shows the neutron count rate profile from the model.
The four panels show how
the profile changes with (a) crater depth, (b) extent of the infill
region, (c) amount of uplift, and (d) crater radius. In all cases
the detector is placed at an altitude of $30$\,km and the effective
beaming of neutrons is taken to be $\alpha=0.5$. The mean altitude for
the LPNS observations being considered at altitudes less than $45$\,km
is $\sim 31$\,km.

Panels (a) and (b) show that, for a $45\,$km radius crater, as the
crater becomes deeper or the 
infill zone smaller, the central peak in count rate over the
crater increases. This happens because these changes enhance the
effect of the neutron beaming seen over the crater. If the beaming of
neutrons is switched off, i.e. $\alpha=0$, and neutrons are allowed to
be reemitted off the crater interior, then the model has
$c(r)/\bar{c}(<~r_{\rm c})=1$ for all $r$.

Panel (c) of Fig.~\ref{fig:mod4} shows how uplifting a $45$\,km crater and
having a constant gradient outer slope that returns to the unperturbed
surface at $x_{\rm e}=2.0\,r_c$ affects the neutron count rate profile. As the
uplift increases, the outer uplifted slope focuses more neutrons onto
the detector when it is over the crater exterior, leading to larger
count rates at $r/r_{\rm c}>1$.

The variation of the model count rate profile with crater radius is
shown in panel (d). Parameters for the model crater shapes are taken
from the fits to the stacked LOLA topographical profiles, as listed in
Table~\ref{tab:modpar}. For the $15$\,km radius craters, the
central peak in the count rate profile occurs on scales too small for
the $45$\,km FWHM of an omnidirectional detector at an altitude
of $30$\,km. Consequently, the profile looks almost flat.
For larger craters, the central peak in neutron count rate becomes
increasingly apparent as the instrumental FWHM corresponds to 
smaller $r/r_c$. The simple geometrical model captures much of the central
bump that is present in the data for the different crater
sizes. However, more apparent is the failure to reproduce the LPNS results at 
$r/r_{\rm c}\gsim 2$, where the model overpredicts the observed count
rate by $\sim 0.3\%$.

The features of the comparison between model and LPNS neutron count
rate profiles are common across the different crater sizes, in both
the deep and shallow craters, and when the observations are split into
high and low altitude subsets and the model is adjusted
accordingly. In all cases, the model appears slightly to underestimate
the count rate observed over the crater. Given that this provides the
normalisation for all count rates, a consequence is that the model
overestimates the normalised count rate at large distances from the crater.

One might wonder if the stacking process, used here to increase the
statistical significance of the measured average neutron count rate profile
features, might introduce systematic effects. For instance, not all craters
in a particular radius range have identical topographical profiles. If
the neutron count rate profile features were especially sensitive to the
deepest craters, which might have only a small impact on the average
topographical profile, then the stacked count rate profile might not reflect
changes in the average topography. However, as the
features in the neutron count rate profiles are small and the
model performs similarly well for subsets of craters selected by
radius or depth, this provides reassurance that such non-linearities are
unimportant here. Consequently, it is evident empirically that the
model based upon the average crater topography does encapsulate the
important features that are responsible for giving rise to the stacked
neutron count rate profile, and the stacking procedure is an
appropriate way to perform this study.

\subsection{Improvements to the simple model}\label{ssec:imp}

\begin{figure}
\begin{center}
\includegraphics[trim=0cm 5cm 1cm 3cm,clip=true,width=0.95\columnwidth]{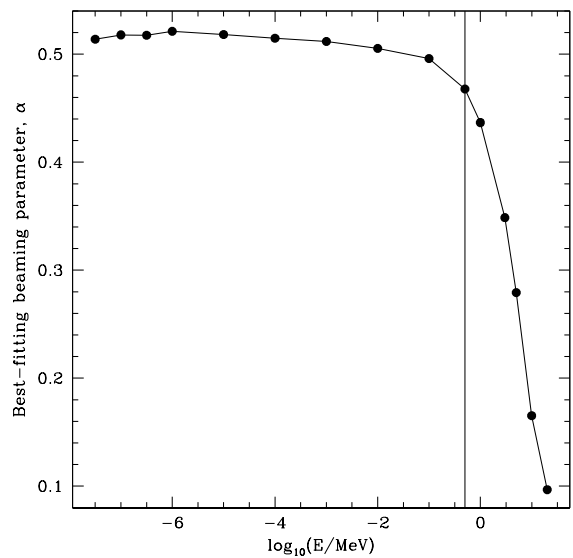}
\end{center}
\caption{Variation of the best-fitting neutron leakage beaming
parameter, $\alpha$, with neutron energy, $E$. The fit to the angular
distribution of leakage neutrons determined using Monte Carlo neutron
transport simulations has the flux as a function of angle
from the surface normal, $\theta$, proportional to
$\cos^{1+\alpha}\theta$. Statistical uncertainties are smaller than
the circles representing the simulation results, and the vertical line
shows the lower limit of $0.5$ MeV for LPNS fast neutrons.
}\label{fig:beam}
\end{figure}

The small difference between the model and LPNS neutron count rate
profiles presumably arises due to an inappropriate assumption in the
simple geometrical model. In this section, the assumptions being made
in the model will be varied to determine what is required in order to
fit the data.

There is no energy dependence in the model
predictions, so the similarity between observed thermal and epithermal
count rate profiles and how they differ from the fast neutron profiles
is immediately suggestive that there is an energy-dependent
misassumption in the model. The assumed beaming factor $\alpha=0.5$ is
relevant for thermal and epithermal neutrons, but for LPNS fast
neutrons, with energies above $0.5$ MeV, the best-fitting $\alpha$
decreases, corresponding to less beaming. Figure~\ref{fig:beam} shows
these results from fits to Monte Carlo neutron transport simulations.
Furthermore, the single parameter power-law fit does not accurately
model the angular distribution of emitted neutrons at the fastest
energies, with the actual distribution being less beamed normal to the
surface. Accounting for the LPNS detector response and the incoming
flux as a function of fast neutron energy suggests that an appropriate
value for $\alpha$ for the model is probably in the range
$0.4-0.45$. This lessening of the beaming acts to suppress the size of
the features in the fast neutron count rate profile, and goes
roughly half way to explaining the difference between the LPNS epithermal and
fast neutron count rate profiles.

Another possible effect that might reduce the fast neutron count rate
profile features is that the
emitted fast neutrons may have an angular distribution that retains some
memory of the direction of the incoming cosmic ray that produced
them. The model assumes that the emitted neutron flux depends only on
the angle from the normal to the surface, and not the azimuthal
angle. Within craters, if fast neutrons are more likely to be emitted
in the forward direction with respect to the incoming cosmic rays, then
this would preferentially aim them into the crater and thus slightly
reduce the count rate measured over the crater. This is qualitatively
consistent with the difference between the fast neutron count rate
profiles and the thermal and epithermal ones. Given these difficulties
in modelling the fast neutron emission, the fast
neutron results will not be considered further.

\begin{figure}
\begin{center}
\includegraphics[trim=0cm 5cm 1cm 3cm,clip=true,width=0.95\columnwidth]{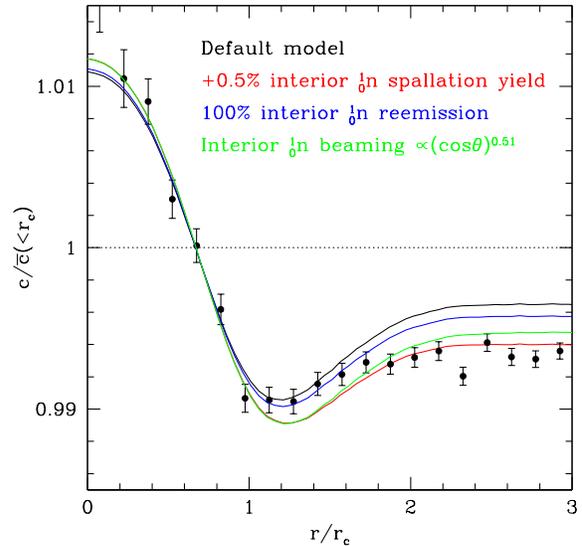}
\end{center}
\caption{
Variation of the model normalised neutron count rate
profile with different changes to the default model for a
$45$\,km-radius crater (black line) relative to the LPNS epithermal
neutron result for the stack of $40<r_{\rm c}/{\rm km}<50$
craters (points). The red curve shows the model resulting from
increasing the emitted epithermal neutron flux per input cosmic ray by
$0.5\%$ within the crater. Including reemission of all neutrons
(rather than the default of none) from the crater interior surface
changes the default model to that shown with a blue curve. The green
line results from increasing the neutron beaming from the surface from
$0.50$ to $0.51$ within the crater only.
}\label{fig:sys}
\end{figure}

Figure~\ref{fig:sys} shows the results found
for the epithermal neutron count rate profiles of $45$ km-radius craters 
when various different model assumptions are made. The common theme in
tweaking the model is the desire to increase the count rate observed
over the crater relative to that observed outside the crater. For
instance, the blue curve results from allowing all neutrons emitted
from within the crater and aimed at another part of the crater
interior to be reemitted rather than absorbed. Details of this
calculation are described in Appendix~(\ref{ssec:cscat}). The
difference between no reemission and complete reemission, which is
presumably also unrealistic, amounts to less than $0.1\%$ in 
the count rate, so is insufficient to make the model fit the data.

Figure~\ref{fig:beam} suggests that $\alpha=0.515$, rather than $0.5$
represents the best description of the beaming of thermal and
epithermal neutrons, but this change is too small to make a
significant difference in the count rate profile. 
Increasing the amount that neutrons are beamed from the surface by
changing $\alpha$ from $0.50$ to $0.51$ within the crater, while
leaving $\alpha=0.50$ for the crater exterior, has a larger impact on
the predicted count rate profile. This is shown by the green curve in
Figure~\ref{fig:sys}, but it still fails to fit
the LPNS results. A similar result is found if the number
of neutrons emitted per incident cosmic ray is increased by $0.5\%$
within the crater only (red curve). While this approximately recovers the
LPNS profile for $r/r_{\rm c}>2$, it predicts a dip at 
$r/r_{\rm c}\sim 1.2$ that is deeper than observed.

\begin{figure}
\begin{center}
\includegraphics[trim=0cm 5cm 1cm 3cm,clip=true,width=0.95\columnwidth]{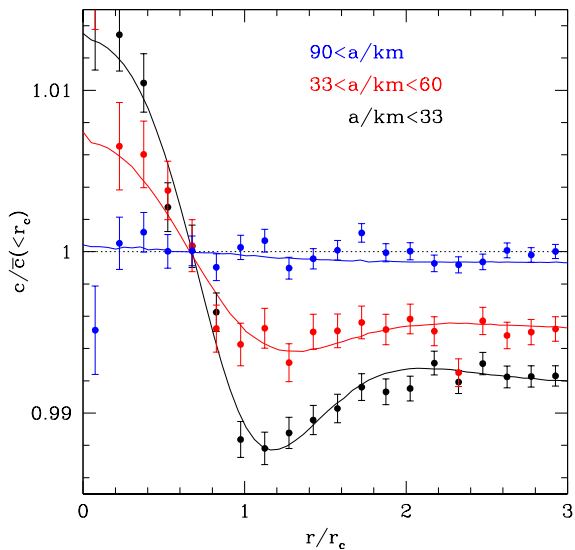}
\end{center}
\caption{
Stacked, normalised radial LP epithermal neutron count rate profiles for 
$40-50$ km radius craters as a function of LP altitude. Blue, black
and red results correspond to detector altitudes of $a<33$ km,
$[33,60]$ km and $a>90$ km respectively. Curves show the corresponding
best-fitting model results.
}\label{fig:neualt}
\end{figure}

No single parameter change that has been considered is able to recover
the observed LPNS neutron count rate profiles. However, a good fit can
be found by including a combination of $75\%$ neutron reemission from the
crater walls and a $0.35\%$ enhancement in the neutron yield for the
region out to $2r_{\rm c}$, which includes the crater interior and
most of the outer uplifted slope. This fits the LPNS count rate
profiles constructed from time series observations taken from
altitudes below $45$ km. The same model also fits the profiles
observed in different altitude ranges, as shown in
Figure~\ref{fig:neualt}. At larger altitudes, the detector footprint
is larger, and this suppresses the amplitude of the features in the
count rate profile, and this behaviour is accurately captured by the model.
This success was not inevitable, but it was
necessary if the tweaks to the model are to be interpreted as telling
us something about the lunar surface. Had Figure~\ref{fig:neualt}
included results from craters/basins with sizes comparable to the
LPNS detector footprint at an altitude of $100$ km, then the
high-altitude data would show a central count rate peak and a drop
outside the crater radius. 

Similar changes to the simple model are able to fit the thermal and
epithermal neutron profiles for all crater size ranges. The required
enhancement is $0.3-0.4\%$ for the crater subsets with $r_{\rm c}<50$
km, with this enhancement ranging out to $1.5-2 r_{\rm c}$ from the
crater centre. 


One possible explanation for the enhanced neutron emission could be
surface or near-surface roughness, of the sort seen by radar measurements
out to twice the crater radius \citep{stickle15}. While previous
neutron transport simulations for planetary surfaces have assumed
emission from a flat surface \citep{law06}, it has been shown that the
neutron leakage flux can be enhanced for non-flat surfaces \citep{druke91}.
If such roughness leads to the increase in emitted neutron flux
required to fit the observed LPNS count rate profiles, then the
neutron count rate could actually be 
sensitive to the physical condition of the lunar surface, making it
complementary to the radar and thermal infrared data sets
\citep{band11,ghent15}. The impact of changing the regolith mass distribution
near the surface can be addressed directly using Monte Carlo neutron
transport simulations that use realistic topographic models of a
planetary surface, as has been done for other planetary bodies \citep{pretty15}. 

\section{Implications for hydrogen in polar cold traps}\label{sec:imp}

The beaming of neutrons increases the count rate measured
by the LPNS when it passes over craters. This is the opposite effect
to that produced by placing hydrogen into permanently shaded regions
(PSRs) within polar craters, which reduces the epithermal count rate. Not 
accounting for the varying topography will lead to underestimates of the
water-equivalent hydrogen cold-trapped into polar PSRs. If the change
in observed count rate due to topography is $\sim 1\%$, then one might
wonder how this could possibly have a significant impact upon the
inferred WEH. However, this small change to the observed count rate is
evident over the entire crater area, whereas the PSR may only cover a
tiny fraction of the crater area. The blurring caused by the response
function of the LPNS can have the effect of levering
a small effect acting over the large crater area into a large effect
in the small PSR area.

\begin{figure}
\begin{center}
\includegraphics[trim=0cm 5cm 1cm 3cm,clip=true,width=0.95\columnwidth]{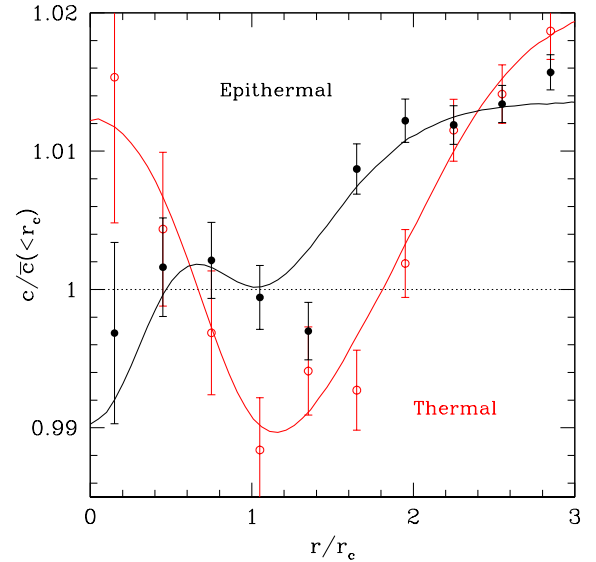}
\end{center}
\caption{
LP epithermal (black filled circles) and thermal
(red open circles) neutron count rate profiles for Cabeus crater.
Curves show models for an $r_{\rm c}=42$\,km crater with the
topography that best-fits that of Cabeus, and
$75\%$ reemission of neutrons hitting the crater interior. The
black curve additionally has a central ``composition'' that emits only
$0.35$ times as many neutrons per cosmic ray as the rest of the surface. The red
curve traces a model where the surface interior to $r_{\rm c}=2.2$
emits $0.97$ times as many neutrons per cosmic ray as the rest of the surface.
}\label{fig:cabeus}
\end{figure}

Rather than considering a general crater, it makes sense to focus on
Cabeus, where LCROSS actually made a local estimate of the WEH weight 
percentage. Cabeus is also significantly deeper than the average crater with a similar
radius. The azimuthally-averaged shape of Cabeus is best fitted with a
profile defined by $d=6.05$\,km, $x_{\rm i}/x_{\rm c}=0.7$, 
$x_{\rm e}/x_{\rm c}=2.3$ and $u=0.85$\,km. Adopting $42$\,km as the
radius of the crater \citep{head10}, and choosing a central circular
disc covering $275$\,km$^2$ as the PSR \citep{luis10}, a model where
the PSR region emits $0.35$ times as many epithermal neutrons per
incoming cosmic ray as the rest of the surface
gives rise to the black line in Figure~\ref{fig:cabeus}.
This reasonably fits the epithermal neutron data for Cabeus, shown
with black filled circles. The thermal neutron data, in contrast, are
well-fitted by a model where the surface interior to $2.2r_{\rm c}$
emits $0.97$ times as many neutrons as the rest of the surface. 

That the thermal and epithermal count rate profiles differ is
consistent with the suggestion that hydrogen in the PSR is responsible
for the odd shape of the epithermal neutron profile, although the
reason for the thermal neutron profile rising above $1$ at $r\gsim
2r_{\rm c}$ is not clear. If one ascribes the lack of an epithermal
central count rate bump entirely to hydrogen in the PSR, then the
factor of $0.35$ in neutron count rate can be converted using the
formula supplied by \cite{law06} into $\sim 4.5$ wt$\%$ WEH. This is a
factor of four greater than was inferred by \cite{luis10}, and 
consistent with the LCROSS results \citep{cola10,stry13}. As Cabeus does not have
a simple crater morphology and the possibility that additional
compositional variation is being suggested by the thermal neutron
profile, this value should be taken with a pinch of
salt. However, it serves to illustrate how the effect of topography
upon remotely-sensed neutron count rates could lead to a significant
bias in the inferred wt$\%$ WEH. Given that many PSRs occupy a larger
fraction of the area of the crater within which they reside, the case
of Cabeus may be a more extreme example of how large an effect
topography can play.

\section{Conclusions}\label{sec:conc}

This study shows that there are some features in the neutron count
rate profiles sensed from orbital detectors as they are flown over lunar
craters located in highland regions. There is a central bump in the
detected count rate, and the mean count rate over the stacked crater is up to
$1\%$ larger than it is outside. This factor is
largest for thermal and epithermal neutrons, but still detectable in
the fast neutrons.

A simple geometrical model has been developed. It predicts
qualitatively very similar behaviour to that observed from the LPNS
thermal and epithermal data sets. The central peak results from the
weak beaming of emitted neutrons normal to the surface \citep{law06},
which is analogous to solar limb darkening. This simple model
underestimates the mean count rate observed over the crater by $\sim 0.3\%$.

To fit the observed stacked count rate profiles well requires a $\sim 0.35\%$
enhancement in the neutron emissivity of the regolith within $\sim
2r_{\rm c}$ of the crater centre. It should be possible, using Monte
Carlo neutron transport simulations, to determine if this can be
achieved by a plausible amount of surface or near-surface roughness. 

The beaming of neutrons over polar craters hosting PSRs may mean that
the concentration of hydrogen in the PSRs has been underestimated in
previous work. For the particular case of Cabeus, where a large crater
contains a relatively small PSR, it was shown that $\sim 4.5$ wt$\%$
WEH within the PSR can reproduce the epithermal neutron count rate
profile, assuming a simple azimuthally-symmetric topographical model
for Cabeus. This is a factor of four times larger than previously
inferred, and is consistent with the value measured using LCROSS
data. In polar craters where the PSR occupies a larger fraction of the
crater, the impact of topography on the inferred wt$\%$ WEH will
be less important.

\begin{acknowledgments}
VRE was supported by the Science and Technology Facilities
Council [ST/L00075X/1].
\end{acknowledgments}

\bibliographystyle{agufull08}

\bibliography{mybib}

\appendix

\section{Details of the model neutron count rate calculation}\label{app:mod}

In order to calculate the model neutron count rate in an orbiting
omni-directional detector, it is easiest to split the surface into a
few distinct regions: the unperturbed surface, the outer uplifted
slope, the crater walls and the flat infill region in the crater
centre. The simpler case with no uplift will be considered first.

\subsection{Neutron count rate from the uncratered surface}\label{ssec:nocrat}

\begin{figure}
\begin{center}
\includegraphics[trim=1cm 7cm 1cm 4cm,clip=true,width=0.95\columnwidth]{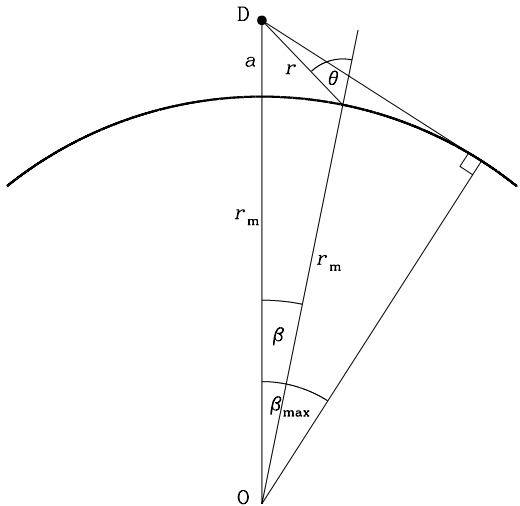}
\end{center}
\caption{
Variables used in the calculation of the count rate from the
unperturbed lunar surface. The origin is placed at the lunar centre,
O, and the detector, D, is at an altitude $a$ above the surface.
}\label{fig:unpert}
\end{figure}

This part of the calculation is very similar to that described in
Appendix B of \cite{pretty06}.
The flux of neutrons a distance $r$ away from a patch of lunar
surface of area d$A$, at an angle $\theta$ to the surface normal, as shown in
Fig.~\ref{fig:unpert}, will satisfy
\begin{equation}
{\rm d }f(r,\theta) \propto \frac{\cos^{1+\alpha}\theta}{r^2}\,{\rm d }A,
\end{equation}
where $\alpha$ represents the effective beaming of neutrons from the
surface. Integrating over $2\pi$ steradians and defining the flux
through the surface as $f_0$, leads to
\begin{equation}
{\rm d }f(r,\theta)=\frac{(2+\alpha)f_0}{2\pi r^2}\cos^{1+\alpha}\theta\,{\rm d }A.
\end{equation}
The flux detected from the whole surface is then
\begin{equation}
f=\frac{(2+\alpha)f_0}{2\pi}\int_0^{2\pi} {\rm d}\phi
\int_0^{\beta_{\rm max}}\left(\frac{r_{\rm m}}{r}\right)^2\cos^{1+\alpha}\theta
\sin\beta~{\rm d}\beta.
\label{general}
\end{equation}
As shown in Fig.~\ref{fig:unpert}, $\beta$ is the angle subtended
at the lunar centre by the vectors to the detector and surface patch
and $\beta_{\rm max}=\cos^{-1}[r_{\rm m}/(r_{\rm m}+a)]$ defines the lunar horizon for
a detector at altitude $a$. Using the sine and cosine rules,
\begin{equation}
\sin\theta=\left(\frac{r_{\rm m}+a}{r}\right)\sin\beta,
\end{equation}
and
\begin{equation}
r^2=(r_{\rm m}+a)^2+r_{\rm m}^2-2r_{\rm m}(r_{\rm m}+a)\cos\beta.
\end{equation}
Defining $t=(r_{\rm m}+a)/r_{\rm m}$ yields
\begin{equation}
f=(2+\alpha)f_0\int_0^{\beta_{\rm max}}\frac{(t\cos\beta-1)^{1+\alpha}\sin\beta
  {\rm d}\beta}{(t^2+1-2t\cos\beta)^\frac{3+\alpha}{2}},
\end{equation}
which can be computed numerically to find the flux from the
uncratered surface.

\begin{figure}
\begin{center}
\includegraphics[trim=1cm 7cm 1cm 7cm,clip=true,width=0.95\columnwidth]{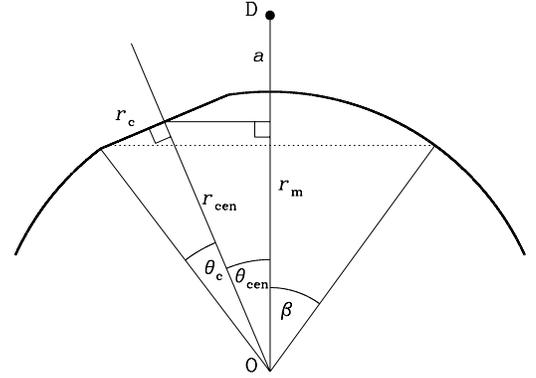}
\end{center}
\caption{
Variables used in the calculation of the count rate from the
unperturbed part of the surface when a crater is present. The dotted
line represents the intersection of the cone with half-opening angle
$\beta$ and the sphere with radius $r_{\rm m}$.
}\label{fig:xsect2}
\end{figure}

The detector has been assumed to be omni-directional in the above
calculation such that the detected count rate is merely proportional
to the flux at the detector. The LPNS is, in fact, cylindrical and
thus is not quite omni-directional. However, comparison of the
inferred 
instrumental point-spread function with that given by \cite{maur04}
shows them to be similar in shape to the extent that correcting for any
differences has a negligible effect upon the results in this paper.

When a crater is inserted into the surface, the integration limits in
equation~(\ref{general}) need to be changed. If the crater centre lies
at the spacecraft nadir, then the minimum $\beta$ is increased so that
the integration starts at the edge of the crater. However, for a more
general crater position it is necessary to find the range of azimuthal
angle $\phi$ that lies outside the crater as a function of $\beta$.
Figure~\ref{fig:xsect2} shows this more general
configuration, where the crater centre subtends an angle
$\theta_{\rm cen}$ at the lunar centre. Without loss of generality, the
detector and crater centre can both be placed in the $x-z$ plane,
where the axes have been chosen such that the y axis is into the page and
the detector is placed on the z 
axis. The required $\phi(\beta)$ can be
found by determining the points where the ring of lunar surface at
$\beta$ intersects with the plane containing the crater rim. Using the
fact that the crater centre lies in the same plane as the crater rim,
one can infer that the rim plane is given by 
\begin{equation}
\underline{x}.\underline{\hat{n}}=r_{\rm cen}=r_{\rm m}\cos\theta_c,
\end{equation}
where the unit normal to the plane is given by
\begin{equation}
\underline{\hat{n}}=\left(
\begin{array}{c}
-\sin\theta_{\rm cen}\\ 
0\\
\cos\theta_{\rm cen}\\
\end{array}
\right).
\end{equation}
Noting the symmetry in the y direction and finding the solution when
\begin{equation}
\underline{x}=r_{\rm m}\left(\begin{array}{c}
\sin\beta\cos\phi\\
\sin\beta\sin\phi\\
\cos\beta\\
\end{array}
\right),
\end{equation}
leads to the following expression for $\phi_{rim}(\beta)$, the angle that
represents the fraction of $\pi$ radians outside the crater for this
$\beta$:
\begin{equation}
\cos\phi_{rim}=\frac{\cos\theta_{\rm cen}\cos\beta-\cos\theta_c}{\sin\theta_{\rm cen}\sin\beta}.
\end{equation}

\subsection{Cosmic ray occlusion within a crater}\label{ssec:occl}

\begin{figure}
\begin{center}
\includegraphics[trim=1cm 6cm 1cm 3cm,clip=true,width=0.95\columnwidth]{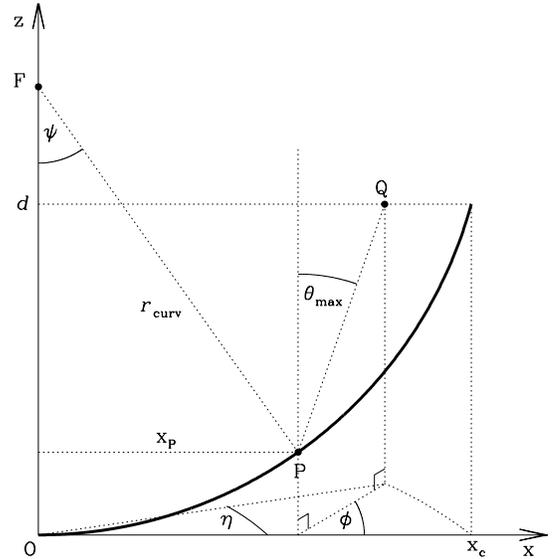}
\end{center}
\caption{
Variables used in the calculation of the sky visibility from a point
P, at $y=0$,
within the spherical cap crater. Points F and Q represent the crater
focus and the point on the crater rim that lies in the vertical plane
through P making an angle $\phi$ with the x
axis. $\theta_{\rm max}(\phi)$ represents the maximum colatitude to
which P sees cosmic rays at an angle $\phi$. $\eta$ is the angle
between the x axis and the point underneath Q in the $x-y$ plane, such
that tan$\eta=y_Q/x_Q$. 
}\label{fig:xsect3}
\end{figure}

The cosmic ray flux impinging upon a unit area of crater interior will
be lower than that incident on the outside, convex surface. Under the
assumption that the cosmic ray flux is isotropic, this is accounted
for by replacing $f_0$ with $f_0\Omega_{\rm i}/\pi$, where
$\Omega_{\rm i}$ is the
cos(incidence angle)-weighted solid angle of visible sky. At a point P
within a spherical cap crater, this is given by
\begin{equation}
\Omega_{\rm i}=\int_0^{2\pi} \int_0^{\theta_{\rm max}(\phi)}\cos
i\sin\theta~{\rm d}\theta {\rm d}\phi,
\label{omega}
\end{equation}
where the incidence angle, $i$, is the angle between the vector
$\underline{\rm PF}$ and the direction $(\theta,\phi)$ and
$\theta_{\rm max}$ is the maximum angle down from the $z$ direction 
that lies above the crater rim, as shown in Figure~\ref{fig:xsect3}.
Q represents the point on the crater rim at this particular
azimuthal angle, $\phi$, and the vector $\underline{\rm PQ}$ makes an
angle $\theta_{\rm max}$ with the z direction. The incidence angle can be
written in terms of the two angular coordinates as
\begin{equation}
\cos i=\frac{(r_{\rm curv}-z_P)\cos\theta-x_P\cos\phi\sin\theta}{r_{\rm curv}},
\label{cosi}
\end{equation}
where $x_P$ and $z_P$ are the $x$ and $z$ coordinates of point P.

Redefining the origin of the coordinate system to be at the base of
the crater, the position of Q is given by
\begin{equation}
\underline{x}_Q=\left(
\begin{array}{c}
x_{c}\cos\eta\\ 
x_{c}\sin\eta\\
d\\
\end{array}
\right),
\end{equation}
with $\eta$ being the angle between the x axis and the point beneath Q
in the $z=0$ plane. Point P has coordinates
\begin{equation}
\underline{x}_P=\left(
\begin{array}{c}
x_P=r_{\rm curv}\sin\psi\\ 
0\\
z_P=r_{\rm curv}(1-\cos\psi)\\
\end{array}
\right),
\end{equation}
where $\psi$ is the angle between the -z direction and 
{\underline{\rm FP}}.
The vertical plane containing P and Q has
\begin{equation}
\underline{\hat{n}}=\left(
\begin{array}{c}
\sin\phi\\ 
-\cos\phi\\
0\\
\end{array}
\right),
\end{equation}
and satisfies
\begin{equation}
\underline{x}.\underline{\hat{n}}=x_P\sin\phi.
\end{equation}
Inserting $\underline{x}_Q$ into this equation yields the following
expression for $\eta$ as a function of $\phi$:
\begin{equation}
\eta=\phi-\sin^{-1}\left(\frac{x_P}{x_{\rm c}}\sin\phi\right).
\label{etaofphi}
\end{equation}
Using the fact that
\begin{equation}
\hat{\underline{\rm PQ}}.\hat{\underline{z}}=\cos\theta_{\rm max},
\end{equation}
one can infer that
\begin{equation}
\cos\theta_{\rm max}=\frac{d-z_P}
{\sqrt{x_{\rm c}^2+x_P^2-2x_{\rm c}x_P\cos\eta+(d-z_P)^2}}.
\label{ctmax}
\end{equation}
For a choice of crater shape and distance from the crater axis,
$x_P$, equations~(\ref{etaofphi}) and~(\ref{ctmax}) determine
$\cos\theta_{\rm max}(\phi)$, which can then be used in conjunction with
equations~(\ref{omega}) and~(\ref{cosi}) to determine the fraction of
sky visible from this point within the crater. 

Extending this approach to the case where there is a flat infilled
region in the crater centre is straightforward. In practice, a table
of $\Omega$ values as a function of $x_P$ is created once, and this is
used, with interpolation, for the two-dimensional numerical integration
to find the flux coming from within the crater.

\subsection{Visibility of a surface patch from the detector}\label{ssec:vis}

Out to the lunar horizon the crater exterior is all visible to the
detector in the case where there is no uplifted rim. However, there
are parts of the crater interior that may not be visible to the
detector. Consequently, it is necessary to see if the line-of-sight
from the detector to the surface patch passes above or below the
crater rim.

Placing the origin of the coordinate system, O, at the lunar centre,
and the detector at
\begin{equation}
\underline{x}_D=(r_{\rm m}+a)\left(
\begin{array}{c}
\sin\theta_{D}\\
0\\
\cos\theta_{D}\\
\end{array}
\right),
\end{equation}
with the z axis going through the crater centre, a general point
within the crater can be written as
\begin{equation}
\underline{x}_P=\left(
\begin{array}{c}
r_{\rm curv}\sin\psi\cos\gamma\\
r_{\rm curv}\sin\psi\sin\gamma\\
r_{\rm m}\cos\theta_c-d+r_{\rm curv}(1-\cos\psi)\\
\end{array}
\right),
\end{equation}
where $\gamma$ represents the angle around from the $x$ axis to point
P. The symmetry of the problem means that the contribution
to the flux coming from $0\le\gamma\le\pi$ is the same as that from
$-\pi\le\gamma\le 0$. Following a similar methodology to that adopted
in Section~\ref{ssec:occl}, the normal to the plane containing O, P
and D can be defined using
$\underline{n}=\underline{x}_P\times(\underline{x}_D-\underline{x}_P)$.
The point Q on the rim determining if the detector is above or below
the crater rim as viewed from P can then be found as the solution to 
$\underline{x}_Q.\underline{\hat{n}}=\underline{x}_P.\underline{\hat{n}}$
with an $x$ coordinate between those of P and D. In this case,
\begin{equation}
\underline{x}_Q=\left(
\begin{array}{c}
x_{c}\cos\eta\\ 
x_{c}\sin\eta\\
r_{\rm m}\cos\theta_c\\
\end{array}
\right)
\end{equation}
and the plane equation is used to determine $\eta$.
For the detector to be able to see point P requires
\begin{equation}
\underline{\hat{z}}.\frac{(\underline{x}_D-\underline{x}_P)}{|\underline{x}_D-\underline{x}_P|} > 
\underline{\hat{z}}.\frac{(\underline{x}_Q-\underline{x}_P)}{|\underline{x}_Q-\underline{x}_P|}.
\end{equation}

These equations, along with those from Sections~\ref{ssec:nocrat}
and~\ref{ssec:occl}, allow the computation of the curves in the top two
panels of Figure~\ref{fig:mod4}. One and two dimensional numerical
integrations are required to evaluate the flux from outside and within
the crater respectively. For the flux from within the crater, it is
also necessary to compute the distance to the detector and the angle
between surface normal and the detector direction, but these are
readily found from the vectors used to determine if the detector can
see that point within the crater. The two dimensional integration to
find the crater flux is simply done over an azimuthal angle ranging
from $0$ to $2\pi$ and the angle from the focus to the crater centre,
$\psi$, running from $0$ to $\psi_{\rm max}=\tan^{-1}[x_{\rm c}/(r_{\rm curv}-d)]$.

\subsection{Uplifted crater rim}\label{ssec:uplift}

Including an uplifted crater rim complicates the calculation
considerably, because parts of the previously unperturbed surface may now
undergo some cosmic ray shadowing and may also no longer be visible
from the detector. Similarly, the outer uplifted slope going from the
crater rim back down to the unperturbed surface is a new topographical
component that also suffers from these issues. In contrast, the
calculation of the flux from the crater itself is only slightly
changed to account for the raising of the entire crater surface.

\subsubsection{Cosmic ray occlusion}

Considering first the occlusion of cosmic rays from the outer uplifted
slope and the unperturbed surface, the symmetry is such that this is
just a function of the distance from the crater centre.
The outer uplifted slope is most conveniently parametrised using an
azimuthal angle, $\epsilon$ and the fraction of the way down the slope
from the rim to the unperturbed surface, $f$. For a point P on the
outer uplifted slope, the
azimuthal variation of the maximum polar angle to which the sky can be
seen, $\theta_{\rm max}(\phi)$, will be set either by the unperturbed
surface or the outer uplifted slope, depending on which is hit first
as the zenith angle increases. 

Choosing the z axis to pass through point P and the crater centre to
lie in the $x-z$ plane at $x<0$, the value of $\theta_{\rm max}$ to the
unperturbed surface is independent of $\phi$. Simple trigonometry gives
\begin{equation}
\cos \theta_{{\rm max},1}=-\sqrt{1-\left(\frac{r_{\rm m}}{r_P}\right)^2},
\end{equation}
with $r_P$ being the
distance of point P from the lunar centre. For sufficiently extended
outer slopes, it is possible for $r_P<r_{\rm m}$, in which case $\cos
\theta_{{\rm max},1}$ is set to $0$.

It may be that the outer uplifted slope itself is the first piece of
lunar surface to intersect the line-of-sight as the zenith angle is
increased at a particular azimuthal angle. In this case,
$\theta_{\rm max}$ is set by the local slope at point P in the azimuthal
direction, $\phi$. For a small displacement on the uplifted slope
having components d$x$ and d$y$, such that $({\rm d}s)^2=({\rm
  d}x)^2+({\rm d}y)^2$ and tan$\phi=\,$d$y/$d$x$, the maximum zenith
angle to the outer uplifted slope can be found from
\begin{equation}
\cos \theta_{{\rm max},2}(\phi)=\sin\left[\tan^{-1}\left(\frac{{\rm d}z}{{\rm d}s}\right)\right].
\end{equation}
\begin{equation}
\frac{{\rm d}z}{{\rm d}s}=\frac{\partial
  z}{\partial x}\frac{{\rm d}x}{{\rm d}s}+\frac{\partial
  z}{\partial y}\frac{{\rm d}y}{{\rm d}s},
\end{equation}
with $\partial z/\partial y=0$, ${\rm d}s/{\rm d}x=1/\cos\phi$ and
$\partial z/\partial x$ being the gradient $g$ from equation~(\ref{grad})
rotated through $\theta_{\rm cen}$ into the coordinate system with P on the z
axis. This leads to
\begin{equation}
\frac{\partial z}{\partial x}=\frac{\sin\theta_{\rm cen}+g\cos\theta_{\rm cen}}{\cos\theta_{\rm cen}-g\sin\theta_{\rm cen}}.
\end{equation}
The value of $\cos\theta_{\rm max}(\phi)$ is taken as the larger of $\cos
\theta_{{\rm max},1}$ and $\cos \theta_{{\rm max},2}(\phi)$, and the cosmic ray
occlusion factor, $\Omega_i/\pi$, at a given fraction of the way down
the outer uplifted slope is calculated using equation~(\ref{omega}).

The cosmic ray occlusion for points on the unperturbed surface, like
that on the outer uplifted slope, is just a function of distance to
the crater centre. It is convenient to place the patch of unperturbed
surface under consideration, P, on the $z$ axis and rotate the crater
centre through an angle $-\theta_{\rm cen}$ about the $y$ axis (moving the
crater 
in the $-x$ direction). Points an angle $\epsilon$ around from the $x$
axis on the crater rim, ${\underline x}_R$,
or the outer edge of the outer uplifted slope, ${\underline x}_E$, can
then be described via
\begin{equation}
\underline{x}_R=\left(
\begin{array}{c}
-(r_{m}\cos\theta_c+u)\sin\theta_{\rm cen}+r_{\rm m}\sin\theta_c\cos\epsilon\cos\theta_{\rm cen}\\ 
r_{m}\sin\theta_c\sin\epsilon\\
(r_{\rm m}\cos\theta_c+u)\cos\theta_{\rm cen}+r_{\rm m}\sin\theta_c\cos\epsilon\sin\theta_{\rm cen}\\
\end{array}
\right)
\label{rim}
\end{equation}
and
\begin{equation}
\underline{x}_E=\left(
\begin{array}{c}
-r_{m}\cos\theta_e\sin\theta_{\rm cen}+r_{\rm m}\sin\theta_e\cos\epsilon\cos\theta_{\rm cen}\\ 
r_{m}\sin\theta_e\sin\epsilon\\
r_{\rm m}\cos\theta_e\cos\theta_{\rm cen}+r_{\rm m}\sin\theta_e\cos\epsilon\sin\theta_{\rm cen}\\
\end{array}
\right)
\label{edge}
\end{equation}
respectively. For a given azimuthal angle $\phi$ around from the $x$
axis as seen from point P on the z axis, it is possible to find any
points on the uplifted outer slope that lie in the plane
\begin{equation}
[\underline{x}_R+f(\underline{x}_E-\underline{x}_R)].\left(
\begin{array}{c}
\sin\phi\\ 
-\cos\phi\\
0\\
\end{array}
\right)=0.
\label{planecr}
\end{equation}
If there are no solutions for $f$ in the range [0,1], then the plane
at fixed $\phi$ does not intersect the uplifted region and
$\cos\theta_{\rm max}(\phi)=0$. If solutions exist, then
equation~(\ref{planecr}) provides a constraint on $f(\epsilon)$ for
points on the 
outer uplifted slope that lie in the plane an azimuthal angle $\phi$
around from the x axis as viewed from point P on the unperturbed surface.
The largest zenith angle from which cosmic rays arrive at point P,
$\theta_{\rm max}(\phi)$, is found using a numerical minimisation algorithm
applied to the set of points on the slope. A root-finding
algorithm is employed to determine $f$ at any given $\epsilon$ as part
of this process. Given $\cos\theta_{\rm max}(\phi)$ as a function of
distance from the crater centre, the cosmic ray occlusion factors can
be found using equation~(\ref{omega}).

\subsubsection{Visibility of surface patch from detector}

Points on either the outer uplifted slope or the unperturbed surface
may not be visible from the detector as a result of the uplifted
region surrounding the crater.

For a point P on the outer uplifted slope to be visible from the
detector, D, the line of sight must not be blocked by either the outer
uplifted slope or the unperturbed surface. If the dot product of the
surface normal at P and the surface-to-detector vector,
$\underline{\Delta}=\underline{x}_D-\underline{x}_P$, is positive,
then P is not blocked by the outer uplifted slope. The unperturbed
surface will block the line-of-sight if the line connecting P
to the detector passes within $r_{\rm m}$ of the lunar centre. Defining the
fractional distance along this line as $v$, such that
$\underline{x}=\underline{x}_P+v\underline{\Delta}$, this happens if 
$0<v_{\rm min}<1$ and $|\underline{x}(v_{\rm min})|<r_{\rm m}$, where $v_{\rm min}$
represents the $v$ for which this line passes nearest to the lunar centre.
With the coordinate system origin at the lunar centre, this leads to 
\begin{equation}
v_{\rm min}=-\frac{\underline{x}_P.\underline{\Delta}}{|\underline{\Delta}|^2}
\end{equation}
and
\begin{equation}
\frac{|\underline{x}(v_{\rm min})|^2}{r_{\rm m}^2}=\frac{|\underline{x}_P|^2-(\underline{x}_P.\underline{\hat{\Delta}})^2}{r_{\rm m}^2},
\end{equation}
where $\underline{\hat{\Delta}}$ is a unit vector in the direction of
$\underline{\Delta}$. These equations allow a quick determination of
whether or not the unperturbed surface blocks the detector's view of a
part of the outer uplifted slope.

To determine the visibility of the unperturbed surface from the
detector, consider placing the detector on the $z$ axis at
$(0,0,r_{\rm m}+a)$ and the crater centre in the $x-z$ plane at $x\le 0$. If
the far point of the edge of the crater outer uplifted slope is
visible above the far point of the rim ($\epsilon=\pi$ in
equations~(\ref{edge}) and ~(\ref{rim}) respectively), then the entire
unperturbed surface is visible from the detector. If this is not the
case, then the plane containing the lunar centre, detector and point P
can be found. The line of intersection of this plane with the uplifted slope
and the minimum zenith angle from P to points on this line follow, and
the visibility is determined by comparison with the zenith angle from
point P to the detector. This is a very similar methodology to that
described to determine the cosmic ray occlusion factor for the
unperturbed surface.

\subsection{Neutron flux impinging upon the crater walls}\label{ssec:cscat}

In the preceding sections of this appendix, the assumption has been
made that any neutrons emitted from within the crater and aimed at the
crater walls are absorbed on contact with the regolith and do not
contribute to the neutron flux emerging from the crater. This is a
simplification, because some of these neutrons will be re-emitted
before being absorbed. The more energetic neutrons may even lead to
nuclear reactions that create more than one lower energy neutron that
escapes from the crater, in which case the crater would be producing a
thermal neutron flux that was an amplified version of
the incident fast neutron flux.

While quantifying the impact of this process requires Monte Carlo neutron
transport simulations, it is possible to use the simple model to
determine how much of the emitted crater flux impinges on the crater
surface as a function of position within the crater.
Following the methodology of the previous sections,
when the crater flux at the detector was determined, it is possible to
place the `detector' on the crater surface and calculate the flux from
the crater that is aimed into the crater surface. The only additional
factor to consider is to include the fact that the normal to the
`detecting' surface is at different angles to the lines-of-sight to
the various other bits of crater surface. Multiplying the detected
flux by the cosine of the incidence angle and integrating over the
entire crater surface leads to the results shown in Fig.~\ref{fig:sys}
for the case where all neutrons are assumed to be reemitted from the surface.

\end{article}
\end{document}